\documentclass[letterpaper]{article} 
\usepackage{aaai2027}  
\usepackage[hyphens]{url}  
\usepackage{graphicx} 
\usepackage{natbib}  
\usepackage{caption} 
\usepackage{algorithm}
\usepackage{algorithmic}
\usepackage{amsmath}

\usepackage{newfloat}
\usepackage{listings}
\DeclareCaptionStyle{ruled}{labelfont=normalfont,labelsep=colon,strut=off} 
\floatstyle{ruled}
\newfloat{listing}{tb}{lst}{}
\floatname{listing}{Listing}

\usepackage{booktabs}
\usepackage{subcaption}
\nocopyright 

\title{SACHA: Semantic-Aware Compression for 3D Gaussian Head Avatars}
\author{
    Zihan Zhang\textsuperscript{\rm 1},
    Shanzhi Yin\textsuperscript{\rm 1},
    Xinju Wu\textsuperscript{\rm 1},
    Bolin Chen\textsuperscript{\rm 2},
    Ru-Ling Liao\textsuperscript{\rm 2},
    Jie Chen\textsuperscript{\rm 2},
    Shiqi Wang\textsuperscript{\rm 1}\corresponding,
    Yan Ye\textsuperscript{\rm 2}
}
\affiliations{
    \textsuperscript{\rm 1}City University of Hong Kong
    \textsuperscript{\rm 2}DAMO Academy, Alibaba Group

}

\begin{document}

\maketitle

\begin{abstract}
Animatable 3D Gaussian head avatars offer high-fidelity and flexible facial rendering, but typically require substantial storage and transmission costs for numerous Gaussian primitives. 
Existing Gaussian head avatar methods overlook the visual saliency of different head semantic regions for more appropriate Gaussian primitive allocation, as well as the efficient compression of trained head avatar sequences. 
To tackle this obstacle, we propose SACHA, a dynamic head avatar compression framework that leverages both semantic-aware density control and appearance-motion decomposition to achieve compact representation and high-quality novel-view rendering of head avatar sequences.
Specifically, the semantic-aware density control guides the adaptive allocation of Gaussian primitives across different head regions with region-adaptive densification and pruning.  
In addition, the appearance-motion decomposed compression further reduces the temporal redundancy of the avatar sequence by transmitting only head-prior parameters for avatar movements.
Together, these designs enable a compact representation for efficient transmission of dynamic Gaussian head avatars while preserving visual fidelity. Experiments demonstrate that SACHA achieves a superior rate-distortion performance over existing Gaussian head avatar representation and compression methods while maintaining high-quality novel-view and novel-expression rendering.

\end{abstract}

\section{Introduction} 

Photorealistic and animatable 3D head avatars have become increasingly important for immersive applications, such as telepresence, virtual reality, remote collaboration, and interactive digital human systems.
To support realistic and interactive user experience, these avatars are expected to provide not only high-fidelity appearance reconstruction but also controllable facial expressions, head motions, and real-time rendering capability. 
Early 3D head avatar methods primarily relied on parametric mesh-based representations~\cite{blanz1999morphable,weise2011realtime} to model facial identity and motion. Later neural-based approaches~\cite{lombardi2018deep} combined explicit geometric proxies with learned dynamics and view-dependent appearance to improve rendering fidelity. 
However, their modeling capabilities are limited by predefined mesh geometry, low-dimensional deformation spaces, and constrained appearance, making it difficult to capture complex geometry and fine appearance details.
Recent advances in neural rendering, from implicit neural representations like Neural Radiance Fields (NeRF)~\cite{mildenhall2021nerf} to explicit representations based on 3D Gaussian Splatting (3DGS)~\cite{kerbl20233d}, have substantially improved the realism and rendering efficiency of 3D head reconstruction.
In addition, recent 3DGS head avatar methods~\cite{giebenhain2024npga, qian2024gaussianavatars,wang2025mega,sun2025svg,wang20253d,lee2026texavatars,song2026progressiveavatars} have integrated head-prior parametric models, such as Faces Learned with an Articulated Model and Expressions~(FLAME)~\cite{FLAMESiggraphAsia2017} and Neural Parametric Head Models~(NPHM)~\cite{giebenhain2023learning}, to facilitate animation of a canonical head representation and enable flexible control over facial dynamics and head poses.

\begin{figure}[t]
    \centering
    \includegraphics[width=1\linewidth]{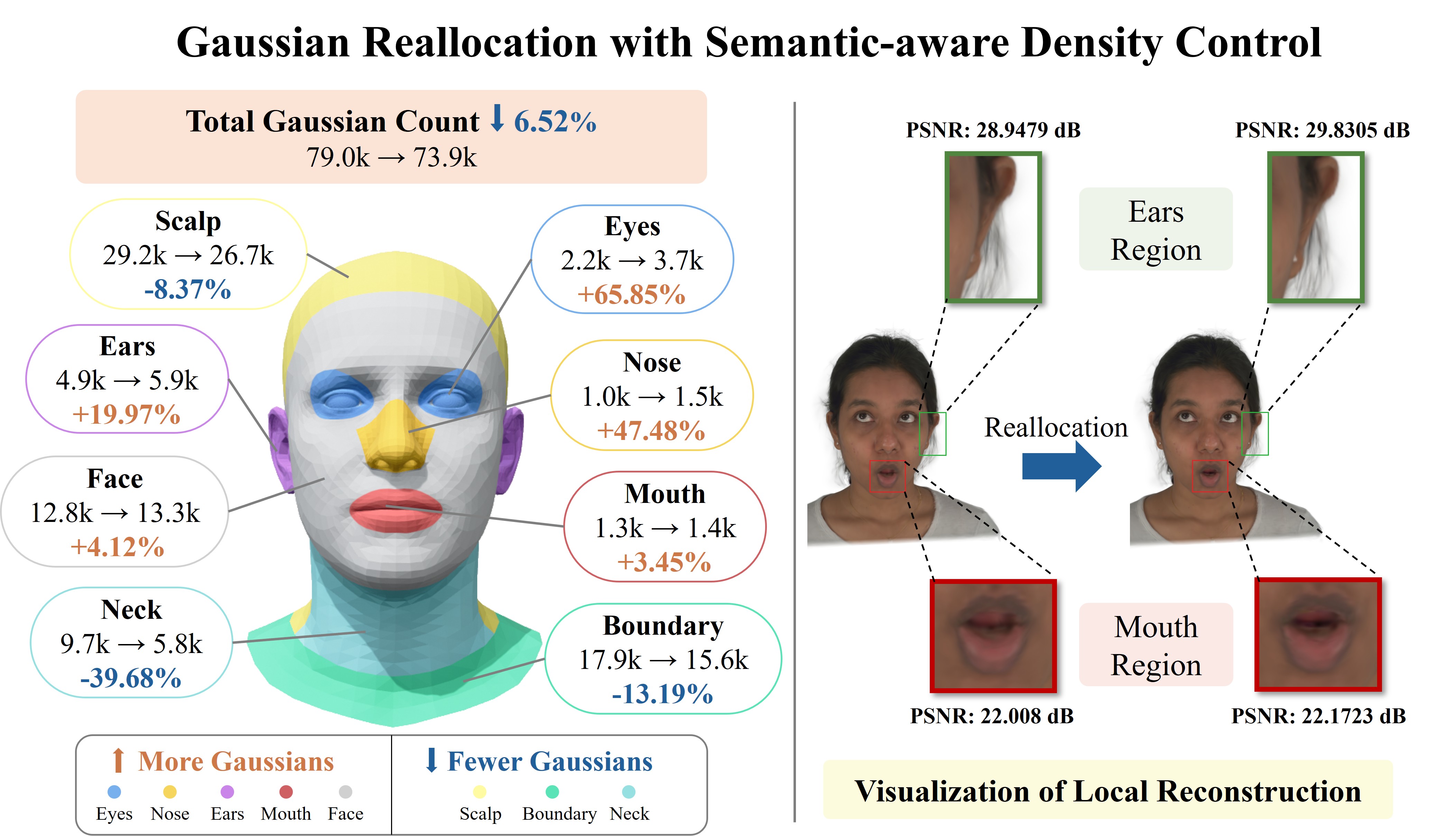}
    \caption{An example of Gaussian reallocation with semantic-aware density control on a specific sequence (Subject 264 from NeRSemble dataset). }  
    \label{fig:intro}
\end{figure}

\begin{figure*}[t]
    \centering
    \includegraphics[width=1\textwidth]{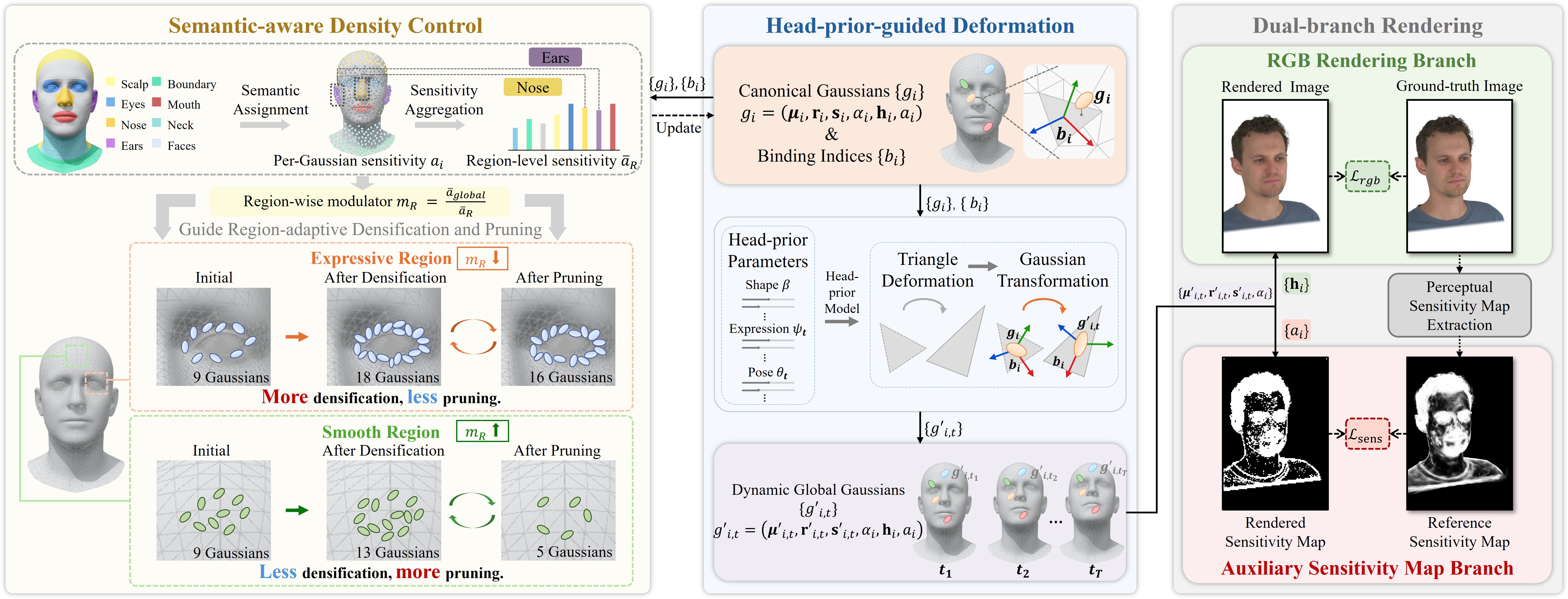}
    \caption{Overview of the semantic-aware canonical Gaussian training pipeline. 
    }
    \label{fig:training}
\end{figure*}

Despite the remarkable progress in head avatar techniques, efficient representation of animatable 3DGS head avatars remains challenging. Preserving fine facial appearance and expression details typically requires a large number of Gaussian primitives, which significantly increases the storage and transmission burden. 
Existing 3DGS compression techniques utilize pruning, quantization, entropy coding, and compact attribute representations~\cite{fan2024lightgaussian, niedermayr2024compressed, liu2024compgs, chen2024hac, tang2026neuralgs} to reduce the representation size for 3DGS scenes while preserving rendering quality. 
These methods are primarily designed for general 3D scenes and typically rely on primitive-level statistics or reconstruction errors to identify redundancy, largely overlooking the structural and semantic priors that are inherent to head avatars.

Specifically, head avatars can be naturally partitioned into multiple semantically meaningful regions, as shown in Figure~\ref{fig:intro}.
These regions differ substantially in visual complexity and deformation intensity.  
Regions such as the eyes and mouth 
exhibit finer appearance details and undergo more significant variations with facial expressions, thus requiring denser Gaussian representations. In contrast, smoother and relatively static regions, such as the neck, 
can be represented with fewer Gaussian primitives.
This motivates a semantic-aware allocation of Gaussian representational capacity rather than global allocation based on a single criterion.
However, most of the current 3DGS head avatar methods directly inherit the density control scheme from the vanilla 3DGS optimization pipeline~\cite{kerbl20233d}, where gradient-based densification and attribute-based pruning are indiscriminately implemented on each primitive. 
This ignores the variation in visual saliency across different head regions, causing potential inefficiency and unnecessary redundancy for 3DGS head avatar compression.

To address these limitations, we propose ``SACHA'', a semantic-aware compression framework for 3D Gaussian head avatars.
The proposed SACHA leverages a sequence-wise canonical Gaussian asset and frame-wise head-prior parameters to build an efficient and animatable framework.
It further learns per-Gaussian perceptual sensitivity through additional supervision and associates the sensitivity with the semantic regions provided by the head-prior model to guide region-adaptive densification and pruning.
Specifically, a region-wise modulator is derived from the learned relative sensitivity of each head region to adjust the densification threshold and pruning ratio during the iterative density control according to regional representation demands. 
To accurately evaluate the perceptual rendering contribution during region-adaptive pruning, both primitive-level and pixel-level sensitivities are considered with a perceptual-score-based Gaussian sorting.
This semantic-aware density control promotes a more effective allocation of Gaussian primitives, as shown in Figure~\ref{fig:intro}.
It encourages refinement in regions with fine visual structures or substantial expression variations while enabling more aggressive Gaussian removal in relatively smooth and less dynamic regions, leading to a favorable rate-distortion~(RD) trade-off for dynamic head avatar coding.
Our main contributions are summarized as follows:
\begin{itemize}
    \item We propose a head-prior-guided semantic-aware compression framework for dynamic Gaussian head avatars that enables appearance-motion decomposition and semantic-aware Gaussian density control.
    \item We develop a region-adaptive densification strategy with a proposed region-wise modulator that aggregates learned perceptual sensitivities within semantic regions and adaptively adjusts region-specific densification thresholds for Gaussian growth.
    \item We design a region-adaptive pruning strategy with a dedicated perceptual importance score for primitive-level Gaussian importance evaluation and further employ the proposed region-wise modulator to adaptively adjust pruning ratios across different semantic regions, thereby enabling region-adaptive and perceptually guided Gaussian pruning.
    \item We conduct comprehensive comparisons with dynamic head avatar representation methods, as well as conventional and learning-based dynamic 3DGS compression methods, and experiment results demonstrate that the proposed method achieves favorable rendering quality with substantially reduced representation sizes.
\end{itemize}

\begin{figure}[t]
    \centering
    \includegraphics[width=0.95\linewidth]{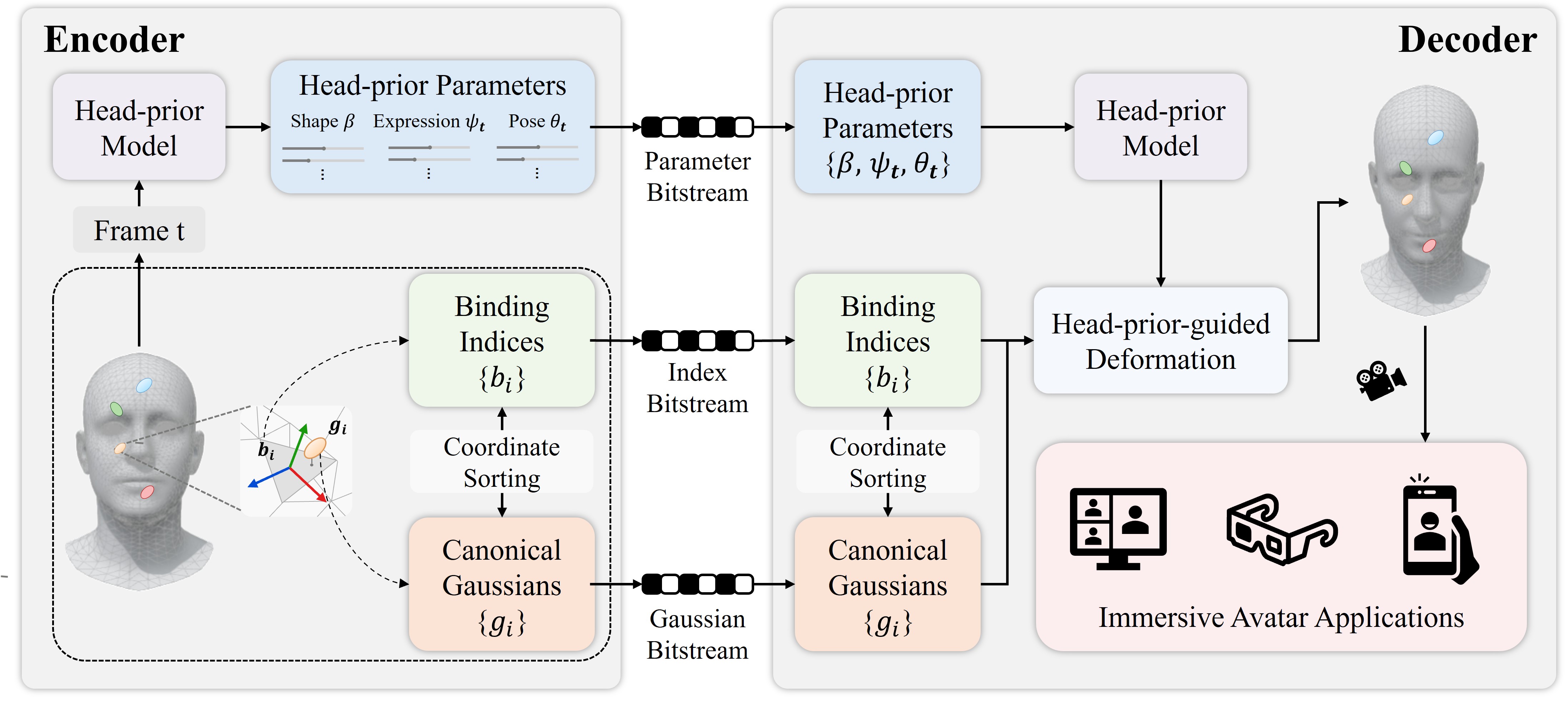}
    \caption{Overview of the proposed appearance-motion decomposed compression framework for 3D Gaussian head avatars.}
    \label{fig:overview}
\end{figure}

\section{Proposed Method}
\subsection{Framework Overview} 
Our proposed SACHA is designed to achieve compact representation and efficient transmission of dynamic Gaussian head avatars. To this end, we factorize each dynamic avatar sequence into two complementary components: a sequence-wise canonical Gaussian representation that preserves the head appearance and identity, and frame-wise head-prior parameters that model temporal motion and expression variations. 
Building on this formulation, the following two subsections describe how we learn compact canonical head Gaussian primitives with semantic-aware density control as detailed in Figure~\ref{fig:training} and how we further perform appearance-motion decomposed compression for efficient transmission and reconstruction as illustrated in Figure~\ref{fig:overview}.

\subsubsection{Semantic-Aware Canonical Gaussian Training.}
We optimize the canonical Gaussians $\mathcal{G}=\{g_i\}_{i=1}^{N}$, where each primitive has geometric attributes, appearance attributes and a learnable perceptual sensitivity attribute, 
\begin{equation}
    g_i=(\boldsymbol{\mu}_i,\mathbf r_i,\mathbf s_i,\alpha_i,\mathbf h_i,a_i).
\end{equation}
The attributes $\boldsymbol{\mu}_i$, $\mathbf r_i$, $\mathbf s_i$, $\alpha_i$, and $\mathbf h_i$ represent the position, rotation, scale, opacity, and spherical harmonic coefficients, respectively, and $a_i$ represents the learned perceptual sensitivity.
During training, these canonical Gaussians are transformed into frame-wise dynamic global Gaussians for rendering through head-prior-guided deformation. 
Specifically, following GaussianAvatars~\cite{qian2024gaussianavatars}, we adopt FLAME~\cite{FLAMESiggraphAsia2017} as the head-prior model and bind each Gaussian $g_i$ to a mesh triangle indexed by $b_i$, with its geometric attributes defined in the local coordinate system of the bound triangle.
For frame $t$, the head-prior parameters deform the mesh, from which the scale factor $k_{b_i,t}$, rotation matrix $\mathbf R_{b_i,t}$, and translation vector $\mathbf T_{b_i,t}$ of the bound triangle are computed to obtain the frame-wise global geometric attributes:
\begin{equation}
\begin{aligned}
\boldsymbol{\mu}_{i,t}^{\prime} &=k_{b_i,t}\mathbf R_{b_i,t}\boldsymbol{\mu}_i +\mathbf T_{b_i,t},\\ 
\mathbf r_{i,t}^{\prime} &=\mathbf R_{b_i,t}\mathbf r_i, \qquad \mathbf s_{i,t}^{\prime}=k_{b_i,t}\mathbf s_i.
\end{aligned}
\end{equation}
Accordingly, the dynamic global Gaussian at frame $t$ is 
\begin{equation}
    g_{i,t}^{\prime}=(\boldsymbol{\mu}_{i,t}^{\prime},\mathbf r_{i,t}^{\prime}, \mathbf s_{i,t}^{\prime},\alpha_i,\mathbf h_i,a_i).
\end{equation}

These dynamic Gaussians are then rendered through a dual-branch rendering scheme comprising an RGB rendering branch and an auxiliary sensitivity map branch~\cite{zhou2025perceptual}, thereby enabling joint optimization of Gaussian geometric attributes, appearance attributes, and per-Gaussian perceptual sensitivities.
Specifically, $\boldsymbol{\mu}_{i,t}^{\prime},\mathbf r_{i,t}^{\prime},\mathbf s_{i,t}^{\prime},\alpha_i,\mathbf h_i$ are used for RGB rendering while $a_i$ is used for sensitivity map rendering, and they share the alpha-compositing weights.  
We aggregate the learned sensitivities of Gaussians within each semantic region to obtain region-level sensitivities that guide region-adaptive densification and pruning during training, yielding a compact set of canonical Gaussians for subsequent sequence compression.

\subsubsection{Appearance-Motion Decomposed Compression.}
After training, the head-prior model can be utilized for 
compact dynamic representation
of the head avatar sequence~\cite{yin2026towards,tang2025hgc}. 
As illustrated in Figure~\ref{fig:overview}, the avatar appearance is represented by canonical Gaussians and only needs to be transmitted once with the binding indices, while temporal variations are driven by per-frame head-prior parameters from the FLAME model. 

At the encoder side, the canonical Gaussians are compressed using an off-the-shelf conventional codec~\cite{gestm}.
To prevent potential mismatches between the decoded canonical Gaussians and their corresponding bound triangles due to the Gaussian re-ordering during the coding process, we establish a deterministic Gaussian order based on the quantized coordinates.
Before encoding, the canonical Gaussians are sorted lexicographically according to their quantized $(x,y,z)$ coordinates, and the corresponding binding indices are rearranged following the same ordering. 
The resulting indices sequence is then remapped using a dictionary and compressed losslessly. 
For head-prior parameters, they are divided into static parameters shared across the sequence and dynamic parameters that vary over frames. Both of them are quantized and encoded using Context-Adaptive Binary Arithmetic Coding (CABAC), where the static components are coded only once and the dynamic components are coded frame by frame.

At the decoder side, the canonical Gaussian attributes, binding indices, and head-prior parameters are decoded from their respective bitstreams. 
The decoded canonical Gaussians are reordered using the same coordinate-based rule and paired with the sequence of decoded indices, thereby restoring the Gaussian-to-triangle correspondence.
For each target frame, the decoded canonical Gaussians are transformed into frame-wise global dynamic Gaussians through head-prior-guided deformation using the recovered bindings and decoded head-prior parameters.
The resulting Gaussians are then rendered to reconstruct the multi-view dynamic head videos.

\subsection{Semantic-Aware Density Control}
To obtain a compact canonical representation, we perform semantic-aware density control during training, adaptively allocating more Gaussians to perceptually important head regions and fewer to less important ones.
Instead of treating all canonical Gaussians uniformly~\cite{kerbl20233d}, we exploit semantic assignments from the head-prior model to distinguish head regions and learned perceptual sensitivities  to guide Gaussian allocation and selection, enabling region-adaptive densification and pruning.

\subsubsection{Region-Adaptive Densification.}
Region-adaptive densification controls Gaussian growth by adjusting the densification threshold for each semantic region.
To define these semantic regions, we divide the FLAME mesh into eight regions according to the official FLAME masks, including the mouth, eyes, nose, ears, boundary, scalp, neck and face, as visualized in Figure~\ref{fig:intro}.
We denote the resulting set of semantic regions by $\mathcal{R}$. 

Since each canonical Gaussian $g_i$ is bound to a triangle of the FLAME mesh indexed by $b_i$, it inherits the semantic label of its bound triangle.  The subset of canonical Gaussians associated with region $R\in\mathcal R$ is then defined as:
\begin{equation}
    \mathcal{G}_R = \left\{   g_i\in\mathcal{G}  \mid \ell(b_i)=R \right\},
\end{equation}
where $\ell(b_i)\in\mathcal{R}$ denotes the semantic label of the FLAME triangle indexed by $b_i$.

For each region $R$, we first compute the region-level perceptual sensitivity as: 
\begin{equation}
    \bar{a}_R = \frac{1}{|\mathcal{G}_R|} \sum_{g_i\in\mathcal{G}_R} a_i,
\end{equation}
where $a_i$ is the learned perceptual sensitivity of Gaussian $g_i$, and $|\mathcal{G}_R|$ denotes the number of Gaussians in region $R$.
Similarly, the global-level perceptual sensitivity is computed as:
\begin{equation}
    \bar{a}_{\mathrm{global}} =
    \frac{1}{|\mathcal{G}|}
    \sum_{g_i\in\mathcal{G}} a_i,
\end{equation}
where $|\mathcal{G}|$ denotes the total number of canonical Gaussians.
Based on the above sensitivity measurements, we introduce a region-wise modulator $m_R$ to adjust the densification threshold: 
\begin{equation}
    m_R = \frac{\bar{a}_{\mathrm{global}}}{\bar{a}_R},
\end{equation}

The threshold $\tau_R$ for region $R$ is then computed as:
\begin{equation}
\tau_R = \begin{cases}
\tau_{\min}, & \tau_0 m_R < \tau_{\min} \\
\tau_{\max}, & \tau_0 m_R > \tau_{\max} \\
\tau_0 m_R, & otherwise,
\end{cases}
\end{equation}
where $\tau_0$ denotes the initial densification threshold shared by all regions, and $\tau_{\min}$ and $\tau_{\max}$ denote the lower and upper predefined bounds of $\tau_R$, respectively.
For regions with sensitivity higher than the global one, $m_R<1$ will be obtained to lower the densification threshold, encouraging Gaussian cloning or splitting. In contrast, less sensitive regions will have $m_R>1$ to prevent unnecessary densification.
The lower and upper bounds prevent excessive threshold modulation and stabilize the region-adaptive densification process.

\subsubsection{Region-Adaptive Pruning.}
In addition to region-adaptive densification,
pruning can further improve compactness by removing redundant primitives while maintaining expressiveness. Following the same intuition, we leverage the region-wise modulator $m_R$ to adapt the pruning ratio for each semantic region, resulting in a region-adaptive pruning strategy.
Specifically, we define the pruning ratio for region $R$ as:
\begin{equation}
\rho_R = \begin{cases}
\rho_{\min}, & \rho_0 m_R < \rho_{\min} \\
\rho_{\max}, & \rho_0 m_R > \rho_{\max} \\
\rho_0 m_R, & otherwise,
\end{cases}
\end{equation}
where $\rho_0$ is a predefined pruning ratio, and $\rho_{\min}$ and
$\rho_{\max}$ denote its lower and upper bounds, respectively.
Regions with larger $m_R$ values are assigned higher pruning ratios for more aggressive pruning, while regions with smaller $m_R$ values receive lower pruning ratios to preserve more primitives.

While the region-specific pruning ratio only determines the number of Gaussians to be removed within each semantic region, it does not indicate which specific primitives to eliminate.
To properly identify the redundant Gaussians in a semantic-aware manner, we design a perceptual importance score that further integrates visual sensitivity into Gaussian importance evaluation.
Building on the importance measure of LightGaussian~\cite{fan2024lightgaussian}, we further incorporate both the reference sensitivity maps $S$ and learned per-Gaussian sensitivities $a_i$, while retaining the accumulated multi-view alpha-blending contribution and the volume-aware factor.

Let $\mathcal{V}$ denote the set of training views and $\Omega_v$ denote the pixel domain of view $v$.
For each view $v\in\mathcal{V}$, let $S_p^{v}$ denote the reference sensitivity value at pixel $p\in\Omega_v$.
The perceptual importance score $q_i$ of Gaussian $g_i$ is defined as:
\begin{equation}
   q_i = \left[ \sum_{v\in\mathcal{V}} \sum_{p\in\Omega_v} w_{i,p}^{v}
   \left(1+\lambda_S S_p^{v}\right) \right] \left(1+\lambda_a a_i\right) \gamma_i,
\end{equation}
where $w_{i,p}^{v}$ denotes the alpha-blending weight of Gaussian $g_i$ at pixel $p$ in view $v$, 
and the volume-aware factor $\gamma_i$ is computed as:
\begin{equation}
\gamma_i =
\left[
\min\left(\frac{V_i}{V_{90}},1\right)
\right]^{0.1},
\qquad
V_i = \frac{4}{3}\pi s_{i,1}s_{i,2}s_{i,3},
\end{equation}
where $(s_{i,1},s_{i,2},s_{i,3})$ are the three components of $\mathbf{s}_i$, and $V_{90}$ denotes the 90th percentile of all Gaussian volumes.
Specifically, the reference sensitivity $S_p^{v}$ reweights the projected contribution of Gaussian $g_i$ according to the local perceptual structure at pixel $p$ in view $v$, while $a_i$ modulates its accumulated multi-view contribution according to the Gaussian-level sensitivity learned during training. 
The coefficients $\lambda_S$ and $\lambda_a$ control the contributions of the pixel-level reference sensitivity and learned Gaussian-level sensitivity, respectively.

With the region-specific pruning ratio and proposed perceptual importance score, 
pruning is performed independently within each semantic region.
For region $R$, $q_i$ ranks Gaussians according to their perceptual contributions, while $\rho_R$ determines how many Gaussians are removed from that region.
Accordingly, the Gaussians in $\mathcal{G}_R$ are sorted in ascending order according to their $q_i$, and the lowest-scoring $\lfloor \rho_R |\mathcal{G}_R| \rfloor$ Gaussians are removed.

Together with region-adaptive densification, this forms a semantic-aware grow-and-prune cycle: region-adaptive densification facilitates refinement in regions with stronger perceptual sensitivity, while region-adaptive pruning preferentially removes low-scoring Gaussians according to region-specific pruning ratios, producing a more compact set of canonical Gaussians while preserving perceptually sensitive facial structures.

\subsection{Optimization}
The 3DGS head avatar is trained by jointly optimizing rendering fidelity, perceptual sensitivity estimation, and Gaussian regularization.
Following GaussianAvatars~\cite{qian2024gaussianavatars}, the rendered RGB images are supervised using a combination of an $\mathcal{L}_1$ term and a D-SSIM term:
\begin{equation}
    \mathcal{L}_{\mathrm{rgb}}
    =
    (1-\lambda)\mathcal{L}_1
    +
    \lambda\mathcal{L}_{\mathrm{D\text{-}SSIM}}.
\end{equation}
We also employ the position and scale regularization terms
$\mathcal{L}_{\mathrm{pos}}$ and $\mathcal{L}_{\mathrm{scale}}$ from GaussianAvatars.
The per-Gaussian sensitivities are learned by supervising the rendered sensitivity map $\hat{S}$ with the reference sensitivity map $S$, which is generated from the ground-truth RGB images following Perceptual-GS~\cite{zhou2025perceptual}.
The sensitivity loss is formulated as:
\begin{equation}
    \mathcal{L}_{\mathrm{sens}}
    =
    \operatorname{BCE}(\hat{S},S).
\end{equation}
The overall training objective is:
\begin{equation}
    \mathcal{L}
    =
    \mathcal{L}_{\mathrm{rgb}}
    +
    \lambda_{\mathrm{sens}}\mathcal{L}_{\mathrm{sens}}
    +
    \lambda_{\mathrm{pos}}\mathcal{L}_{\mathrm{pos}}
    +
    \lambda_{\mathrm{scale}}\mathcal{L}_{\mathrm{scale}}.
\end{equation}

The optimization consists of a warm-up stage followed by a region-adaptive density control stage.
During warm-up, the modulators $m_R$ are fixed at 1, such that all regions use the same densification threshold, and region-adaptive pruning is disabled.
This stage stabilizes the optimization of the canonical Gaussian attributes and per-Gaussian sensitivities.
After warm-up, the region-level perceptual sensitivities $\bar{a}_R$ are periodically recomputed and used to update the modulator $m_R$, region-adaptive densification thresholds $\tau_R$, and region-adaptive pruning ratios $\rho_R$.
Region-adaptive densification is then performed at regular intervals, while region-adaptive pruning is periodically applied according to the perceptual importance score $q_i$ and the corresponding region-specific pruning ratios $\rho_R$.

\begin{figure*}[t]
    \centering
    \includegraphics[width=0.85\textwidth]{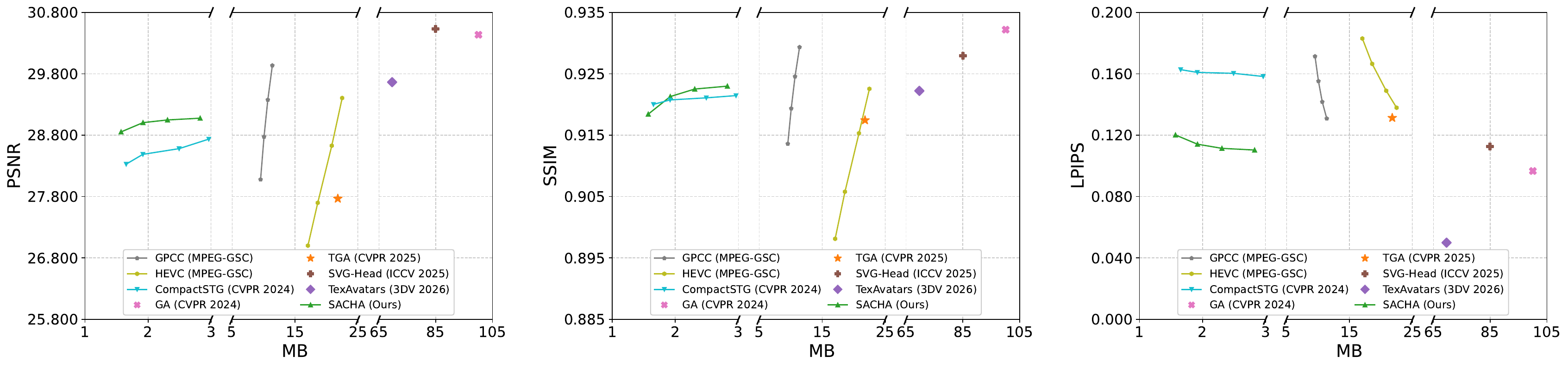}
    \caption{Rate-distortion performance comparisons for novel-view synthesis in terms of PSNR, SSIM, and LPIPS.}
    \label{fig:nvs_rd}
\end{figure*}

\section{Experimental Results}
\subsection{Experimental Settings}
\subsubsection{Datasets.}
We conduct experiments on the NeRSemble dataset~\cite{kirschstein2023nersemble}, which provides synchronized multi-view facial videos captured from 16 calibrated viewpoints. 
10 subjects are selected (074, 104, 165, 218, 253, 264, 302, 304, 306, and 460) in the experiments. 
All images are downsampled to a resolution of $802 \times 550$ pixels.
We consider two evaluation settings: novel-view synthesis and novel-expression synthesis. 
For each subject, the model is trained using 15 of the 16 viewpoints, with one complete expression sequence additionally excluded from all training viewpoints.
For novel-view synthesis, we evaluate 100 frames per subject from the held-out viewpoint, which cover diverse facial expressions across the 10 subjects.
For novel-expression synthesis, we evaluate 100 frames per subject from the held-out expression sequence across all 16 viewpoints, where different held-out expressions are selected for different subjects to cover diverse unseen expressions.


%

\subsubsection{Implementation Details.}
The proposed method is implemented in PyTorch, and all experiments are conducted on a single NVIDIA GeForce RTX 4090D GPU. 
For each subject, the model is independently optimized for 360,000 iterations using the Adam optimizer. 
Unless otherwise specified, the Gaussian representation, FLAME binding, rendering settings, and reconstruction-related loss weights follow GaussianAvatars~\cite{qian2024gaussianavatars}.
The optimization starts with a warm-up stage, where the first 110,000 iterations are performed using global density control. 
Then, the proposed region-adaptive densification and pruning strategies are activated. 
During this stage, the region-level sensitivities $\bar{a}_R$ and global-level sensitivities $\bar{a}_{global}$ are recalibrated every $T_{\mathrm{calib}}=50{,}000$ iterations, followed by the recalculation of region-wise modulator $m_{R}$, region-specific densification thresholds $\tau_R$, and pruning ratios $\rho_R$.
The base densification threshold and pruning ratio are set to $\tau_0=3\times10^{-4}$ and $\rho_0=0.2$, respectively. 
The sensitivity loss weight and the learning rate of the perceptual sensitivity parameters are set to $1\times10^{-2}$ and $1\times10^{-4}$, respectively.

After training, the canonical Gaussian asset is compressed using GeS-TM~\cite{gestm}  with four compression levels. 
The reported representation size includes all data required for reconstruction, including the compressed canonical Gaussian asset, Gaussian-to-triangle binding indices, and FLAME parameters, and is measured in megabytes (MB). 
We evaluate reconstruction quality using Peak Signal-to-Noise Ratio (PSNR), Structural Similarity Index Measure (SSIM)~\cite{wang2004image}, and Learned Perceptual Image Patch Similarity (LPIPS)~\cite{zhang2018unreasonable}.
We also use the RD curves to compare the compression performances.

\begin{figure*}[t]
    \centering
    \includegraphics[width=0.85\textwidth]{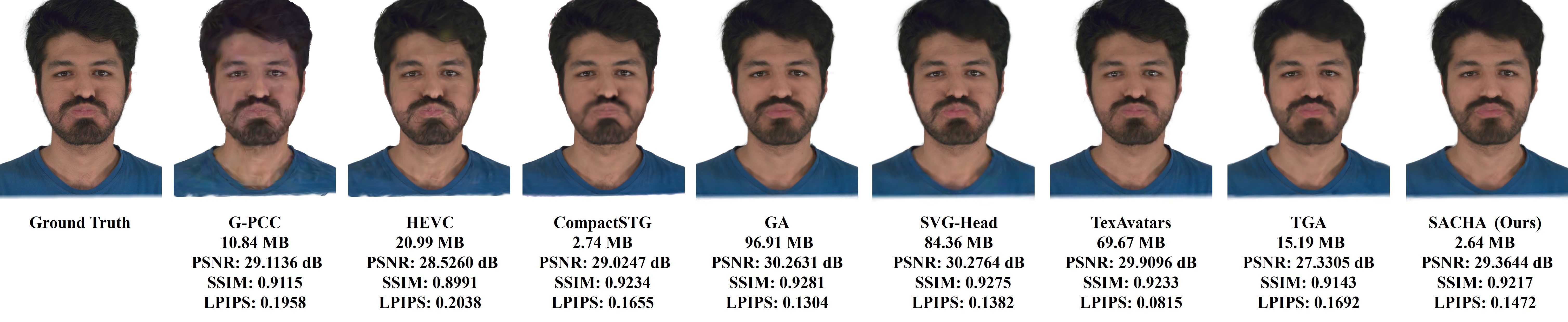}
    \par
    \centerline{\small (a) Subject 074}
    \medskip
    \includegraphics[width=0.85\textwidth]{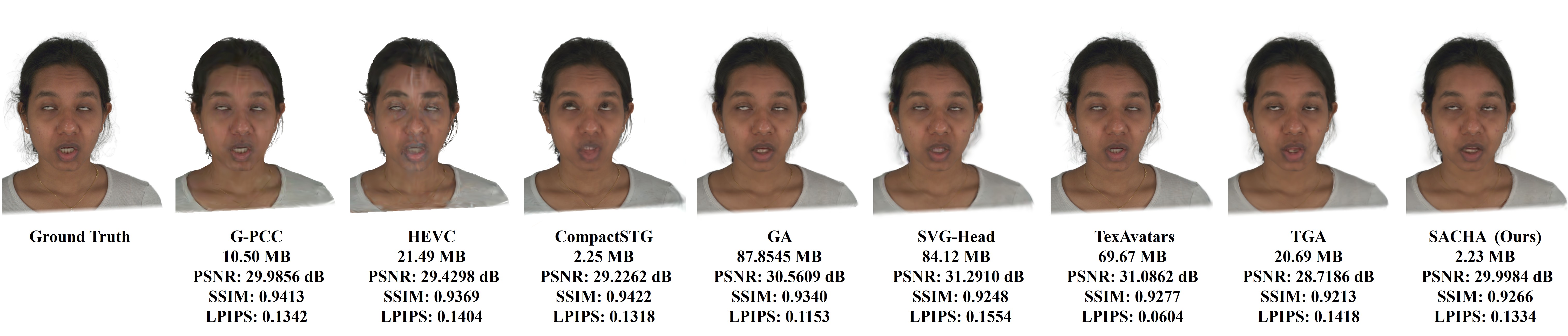}
    \par
    \centerline{\small (b) Subject 264}
    \caption{Subjective comparisons with G-PCC, HEVC, CompactSTG, GA, SVG-Head, TexAvatars, and TGA on subjects 074 and 264 from the NeRSemble dataset.}
    \label{fig:subjective_nvs}
\end{figure*}

\subsubsection{Comparison Methods.}
We compare the proposed method with four representative dynamic head avatar methods, including GaussianAvatars (GA)~\cite{qian2024gaussianavatars}, TensorialGA~ (TGA)~\cite{wang20253d}, TexAvatars~\cite{lee2026texavatars}, and SVG-Head~\cite{sun2025svg}.
Since these methods do not support variable-rate compression, they are reported at their respective native operating points.

In addition, we consider three compression-oriented methods.
Two of them are selected from Joint Exploration Experiment 6.2 between WG 4, WG 5 and WG 7 of the Moving Picture Experts Group (MPEG)~\cite{mpeg}, including a point-cloud-compression~(PCC)-based method with G-PCC~\cite{gpcc} codec and a video-based method~\cite{video-based} with High Efficiency Video Coding~(HEVC)~\cite{sullivan2012overview}. 
We also compare the proposed method with CompactSTG~\cite{lee2024compact}, a learning-based dynamic 3DGS compression method.
For the G-PCC- and HEVC-based methods, we train frame-by-frame sequences by independently optimizing each frame as a static 3DGS representation for 2,000 iterations and generate multiple rate points by varying the corresponding quantization parameters.
For CompactSTG, the entire sequence is modeled as a dynamic scene and optimized for 25,000 iterations with different mask-loss weights used to control the pruning ratio.

For novel-view synthesis, we compare with all applicable methods on the same test frames, while novel-expression comparisons are limited to four dynamic head avatar methods that support expression-driven synthesis.

\subsection{Performance Comparisons}
\subsubsection{Rate-Distortion Performances of Novel-View Synthesis.}
The RD performance of the proposed method compared with the comparison methods in terms of PSNR, SSIM, and LPIPS for novel-view synthesis is shown in Figure~\ref{fig:nvs_rd}. 
It can be observed that the proposed method achieves favorable RD performance across the three evaluation metrics, particularly within the low bit-rate under 5 MB.
Specifically, CompactSTG operates within a similar bit-rate range, but its RD performance is less competitive than that of the proposed method, especially in terms of PSNR and LPIPS. 
The G-PCC- and HEVC-based methods achieve a higher reconstruction quality at their highest-rate points because each frame is independently optimized.
However, such frame-by-frame representations require substantially larger representation sizes around 10 MB, and their reconstruction quality degrades significantly at lower rates.

In contrast, the proposed method reduces temporal redundancy through the shared canonical Gaussian asset and compact head-prior parameters, and further reduces the redundancy of the canonical asset through semantic-aware density control, thereby achieving better reconstruction quality at much lower rates. 
For compared head avatar representation methods, although some of them achieve higher reconstruction quality, they require considerably larger representation sizes more than 20 MB. 
\subsubsection{Subjective Comparisons of Novel-View Synthesis.}
The subjective comparisons of proposed method and comparison methods are shown in Figure~\ref{fig:subjective_nvs}. 
With similar objective quality, the proposed method requires the smallest representation size while producing more visually pleasing results.
Specifically, CompactSTG tends to over-smooth fine facial structures, whereas G-PCC and HEVC introduce more noticeable distortions and color inconsistencies around the facial features, hair, and head boundaries.
In contrast, the proposed method better preserves facial appearance and local details at a substantially lower rate.
For GA, SVG-Head, and TexAvatars, they achieve similar subjective quality as the proposed method
but require considerably larger representation sizes, while TGA is more compact but exhibits more noticeable over-smoothing around facial features.

\subsubsection{Novel-Expression Synthesis.}
Table~\ref{tab:novel_expression} presents the novel-expression synthesis performance comparison with dynamic head avatar methods.
The proposed method is evaluated using its highest-quality compression setting.
Despite its substantially smaller representation size, our method achieves competitive reconstruction quality compared with existing head avatar representations.
Specifically, it outperforms SVG-Head in terms of PSNR, SSIM, and LPIPS, while achieving results comparable to GA and TGA. 
TexAvatars obtains the best PSNR and LPIPS among the compared methods but under-performs the proposed method in terms of SSIM. 
These results indicate that the proposed method generalizes well to unseen facial expressions and maintains competitive reconstruction quality across different viewpoints.

\begin{table}[t]
\centering
\small
\setlength{\tabcolsep}{2pt}
\renewcommand{\arraystretch}{1.0}
\begin{tabular}{lcccc}
\toprule
Method & Size (MB)& PSNR$\uparrow$ & SSIM$\uparrow$ & LPIPS$\downarrow$ \\
\midrule
GA~\cite{qian2024gaussianavatars} & 100.08 & 25.5097 & 0.9102 & 0.1103 \\
TGA~\cite{wang20253d}   & 21.78  & 25.8089 & 0.9125 & 0.1364\\
TexAvatars~\cite{lee2026texavatars}    & 69.67  & 26.5422 & 0.9087 & 0.0596 \\ 
SVG-Head~\cite{sun2025svg}      & 84.99  & 24.9033 & 0.9010 & 0.1270\\
\textbf{SACHA~(Ours)}       & \textbf{2.83}  & \textbf{25.4799
}& \textbf{0.9103} & \textbf{0.1202} \\
\bottomrule
\end{tabular}
\caption{Quantitative comparisons for novel-expression synthesis in terms of PSNR, SSIM, and LPIPS.} 
\label{tab:novel_expression}
\end{table}

\subsection{Ablation Study}

We conduct ablation experiments to evaluate the effectiveness of region-adaptive densification, the proposed perceptual importance score, and region-adaptive pruning. 
All ablation variants and their corresponding RD results are shown in
Figure~\ref{fig:ablation_rd}.
\textit{Default} denotes the vanilla 3DGS configuration, which uses a single global densification threshold and opacity-based pruning. 
\textit{RAD} and \textit{RAP} denote region-adaptive densification and region-adaptive pruning, respectively. \textit{GP} denotes global pruning with a fixed pruning ratio for all regions, while \textit{LGS} and \textit{PIS} denote the pruning score proposed in LightGaussian~\cite{fan2024lightgaussian} and the perceptual importance score proposed in our method, respectively.


\subsubsection{Effectiveness of Region-Adaptive Densification.}
To evaluate the effectiveness of region-adaptive densification, we compare \textit{Default} and \textit{RAD} under vanilla 3DGS pruning strategy.
\textit{Default} applies a single densification threshold to all Gaussian primitives, whereas \textit{RAD} adaptively adjusts the region-specific thresholds according to region-level perceptual sensitivity.
As shown in Figure~\ref{fig:ablation_rd}, \textit{RAD} achieves better RD performance than \textit{Default}, demonstrating that region-adaptive densification enables a more effective allocation of Gaussian primitives.

\subsubsection{Effectiveness of the Proposed Perceptual Importance Score.}
To evaluate the effectiveness of proposed perceptual importance score, we compare the fixed-ratio pruning with importance score proposed in LightGaussian~\cite{fan2024lightgaussian} and our proposed perceptual importance score, denoted as \textit{RAD + GP w/ LGS} and \textit{RAD + GP w/ PIS}, respectively.
As shown in Figure~\ref{fig:ablation_rd}, 
at comparable rates, the proposed perceptual importance score brings clear improvements in PSNR and SSIM with the same pruning ratio.
These results show that incorporating both pixel-level reference sensitivity and learned Gaussian-level sensitivity provides a more reliable ranking of primitives for pruning.

\subsubsection{Effectiveness of Region-Adaptive Pruning.}

\begin{figure}[t]
\centering
\includegraphics[width=\linewidth]{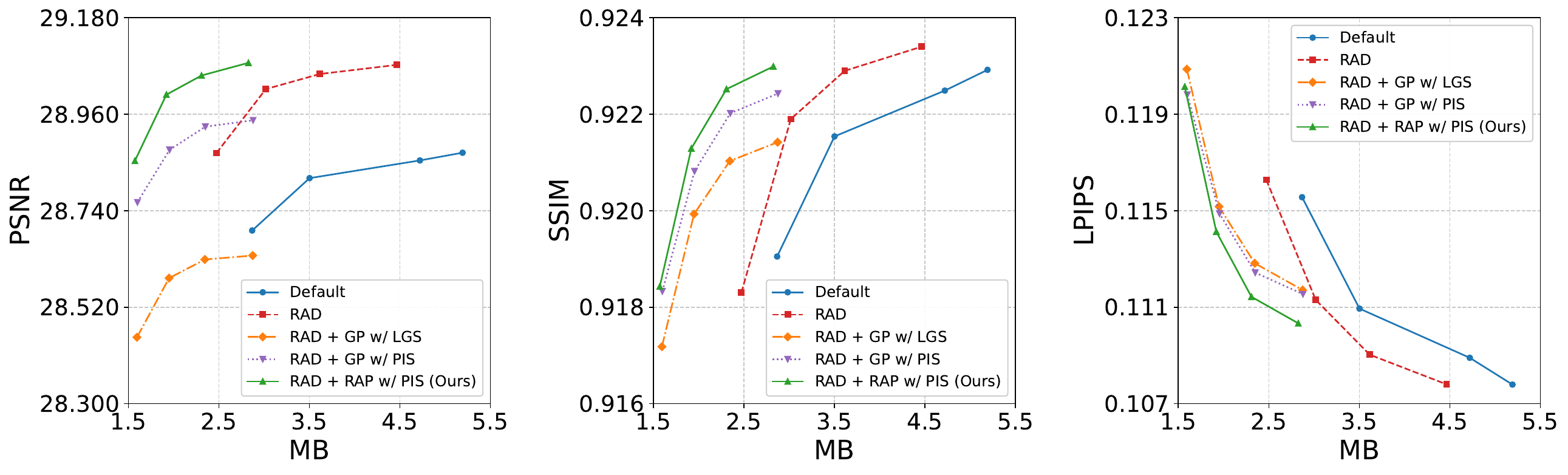}
\caption{Rate-distortion performance comparisons for novel-view synthesis of the ablation variants in terms of PSNR, SSIM, and LPIPS.}
\label{fig:ablation_rd}
\end{figure}

Finally, we evaluate the effectiveness of region-adaptive pruning by comparing it to fixed-ratio pruning with perceptual importance score, denoted as \textit{RAD + GP w/ PIS} and \textit{RAD + RAP w/ PIS}, respectively.
As shown in Figure~\ref{fig:ablation_rd}, \textit{RAD + RAP w/ PIS} achieves a more favorable RD performance than \textit{RAD + GP w/ PIS} across all three evaluation metrics. 
These results demonstrate that, by applying a distinct pruning ratio for each semantic region, a more appropriate Gaussian removal can be achieved across different head regions.

\subsubsection{Regional Qualitative Analysis.}
We compare \textit{Default + GP w/ PIS} with \textit{RAD + RAP w/ PIS} to assess how semantic-aware density control affects Gaussian allocation and local reconstruction quality.
Both variants use the proposed perceptual importance score for pruning and differ only in whether semantic regions are used to adapt the densification thresholds and pruning ratios. 
For Subject 264, Figure~\ref{fig:intro}  shows that semantic-aware density control allocates more Gaussians to regions with fine visual structures while reducing redundant primitives in smoother regions.
This reallocation reduces the total Gaussian count while improving local reconstruction quality in both the ear and mouth regions.

\section{Conclusion}
In this paper, we propose SACHA, a semantic-aware compression framework for animatable 3D Gaussian head avatars. 
The proposed framework decomposes dynamic head avatars into a compact canonical Gaussian representation and frame-wise head-prior parameters, reducing temporal redundancy while maintaining flexible animation capability. 
To further improve the compactness of the canonical representation, we develop a semantic-aware density control strategy that leverages head-prior semantics and learned perceptual sensitivity to  jointly perform region-adaptive densification and pruning.
The proposed SACHA enables more efficient allocation of Gaussian primitives across different semantic regions, and
experimental results demonstrate that SACHA achieves favorable rate-distortion performance, 
while preserving high-quality multi-view rendering, 
showing its potential for efficient transmission of dynamic Gaussian head avatars.

\bibliography{aaai2027}


\end{document}